\documentclass{article}
\usepackage{paithan}
\usepackage{tikz}
\usetikzlibrary{shapes, positioning, backgrounds, fit}
\usepackage{ifxetex}
\ifxetex
  \usepackage{fontspec}
\else
  \usepackage[T1]{fontenc}
  \usepackage[utf8]{inputenc}
  \usepackage{lmodern}
\fi
\usepackage[utf8]{inputenc}
\usepackage[english]{babel}

\title{Computational Properties of \ruleset{Slime Trail}}
\author{Matthew Ferland and Kyle Burke}

\begin{document}
\maketitle

\tikzstyle{empty} = [draw,ultra thick,circle,minimum size = 1cm]
\tikzstyle{red} = [draw,ultra thick,circle,minimum size = 1cm, fill=red]
\tikzstyle{blue} = [draw,ultra thick,circle,minimum size = 1cm, fill=blue]
\tikzstyle{cLine} = [draw, ultra thick]
\tikzstyle{cDashed} = [draw, thick, dashed]
\tikzstyle{cPointed} = [draw, ultra thick, ->]

\begin{abstract}
We investigate the combinatorial game \ruleset{Slime Trail}. This game is played on a graph with a starting piece in a node. Each player's objective is to reach one of their own goal nodes. Every turn the current player moves the piece and deletes the node they came from. We show that the game is PSPACE-complete when played on a planar graph.
\end{abstract}

\section{Introduction}

\subsection{Combinatorial Game Theory}
A combinatorial game is a game without chance or hidden information, with alternating turns between two players. There is a verbose mathematical system to analyze these types of games created by Richard Guy, John Conway, and Elwyn Berlekamp in 1982\cite{WinningWays:2001}. This has sparked a field of mathematics dedicated to formally studying these games.

\subsection{Algorithmic Combinatorial Game Theory}
Algorithmic combinatorial game theory is the study of combinatorial games from a computer science perspective. Of particular interest is determining the computational tractability of solving individual games \cite{AlgGameTheory_GONC3}.

\subsection{Rules of \ruleset{Slime Trail}}
The exact rules for \ruleset{Slime Trail} sometimes vary. The ruleset that the creator Bill Taylor used\footnote{\url{http://homepages.di.fc.ul.pt/~jpn/gv/slimetrail.htm}} is different from the ruleset that the Portugese group Ludus uses\footnote{\url{http://ludicum.org/cnjm/2016-2017-cnjm13/regras-dos-jogos-do-cnjm13/view}}, for example. For the purposes of this paper, we will define a generalized ruleset for the game.

\theoremstyle{definition}
\begin{definition}[\ruleset{Slime Trail}]
\ruleset{Slime Trail} is played on a connected graph, with at least one vertex colored blue, at least one other colored Red, and one vertex with a moveable piece or token.  The two players alternate turns moving the token to an adjacent node, then marking the previous space (where the token was) a third color (usually green).  The token can never be moved back to one of these "slimed" spaces.

A player wins when the token is moved onto a space of their color.  Since we want to make sure that one player can still win, the token is not allowed to move the token to a space where it can't reach at least one goal vertex. However, it is possible to move so that none of a player's goal nodes are reachable. In this case, the other player automatically wins.
\end{definition}

\subsection{History}
\ruleset{Slime Trail} was created by Bill Taylor in 1992. Since it's creation, it has been widely played. It was analyzed by Dave Boll in 1993\footnote{\url{http://www.gamecabinet.com/rules/Slimetrail.html}}. It was used to study blind mathematical play during the International Council for Children’s Play 26th World Play Conference in 2012\footnote{\url{http://www.iccp-play.org/documents/tallinn/dias.pdf}}. It continues to be used by Ludus every year for their popular annual mathematical game tournaments, which started including \ruleset{Slime Trail} in 2008\footnote{\url{http://ludicum.org/}}. In the 2016/2017 competition, 1500 students competed in the finals\footnote{\url{http://www.di.fc.ul.pt/~jpn/cnjm13/index.htm}}. The popularity of the game provides motivation for proving it to be computationally intractable.



\section{Computational Complexity}
\label{sec: complexity}

\subsection{Main theorem}
\textbf{Theorem 3.1} \textit{\ruleset{Slime Trail} is PSPACE-complete when played on a planar graph}

Theorem 3.1 is true if and only if \ruleset{Slime Trail} is both PSPACE-hard and in PSPACE. We show it is in PSPACE in \cref{Lemma 3.2}, and spend the remainder of the section proving the game to be PSPACE-hard.

\begin{lemma}[\ruleset{Slime Trail} is in PSPACE]
\label{Lemma 3.2}
Since the number of plays is no more than the number of nodes in the game board minus one, the depth of every branch of the game tree is linear. Thus, in a polynomial amount of space we can determine the result of following one path of the game tree. In order to search for a winning result, we can systematically try each possible individual game branch. Therefore, we only need enough space enough to evaluate one branch at a time, and to keep track of which branches we have already visited.

This will require only $O(m^2)$ space, where $m$ is the number of nodes on the board. Therefore, in polynomial space, we can evaluate all the possible outcomes of the game tree until we either find a winning strategy or determine that none exists.
\end{lemma}

\subsection{Overview of the QBF reduction}

We are able to show that, when played on a planar graph, \ruleset{Slime Trail} is PSPACE-hard. We can do this by reducing the quantified Boolean formula problem, or QBF, to \ruleset{Slime Trail}\cite{PapadimitriouBook:1994}. The QBF is the problem of determining whether the Boolean formula $\exists x_1 : \forall x_2 : \exists x_3 : \forall x_4 : \ldots Q_n x_n  \phi(x_1, x_2, \ldots x_n)$ is true. In this notation, $\phi(x_1, x_2, \ldots x_n)$ is a conjunctive normal form formula using the literals $x_1$ through $x_n$ while $Q_n$ is a quantifier (either a $\exists$ or $\forall$). This problem is known to be PSPACE-complete\cite{DBLP:journals/ipl/AspvallPT79}.

Because of the inherent alternation in the QBF, it is often used to demonstrate that two player games with non-obvious strategies are PSPACE-hard\cite{PapadimitriouBook:1994}. We can see that fulfilling a QBF is much like playing a game. The first player chooses a variable for $x_1$. Next, the second player chooses a variable for $x_2$. Then, the first player chooses a variable for $x_3$, and so on.

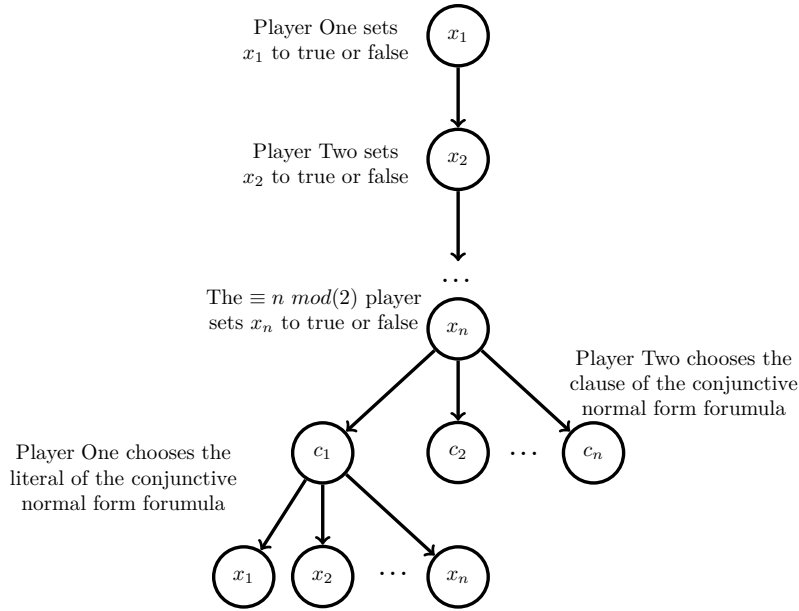
\begin{figure}[h!]
\begin{centering}
\scalebox{.8}{%
\begin{tikzpicture}[scale=.75, every node/.style={draw,ultra thick,circle,minimum size = .3cm} ]
    \node (x_1) [empty, label={[align=center, shift={(-2.2,-2.2)}]Player One sets \\$x_1$ to true or false}] {$x_1$};
    \node (x_2) [empty, label={[align=center, shift={(-2.2,-2.2)}]Player Two sets \\$x_2$ to true or false}, below=of x_1]             {$x_2$};
    \node (x_n) [empty, label={[align=center, shift={(-2.4,-2.2)}]The $\equiv n \ mod(2)$ player \\ sets $x_n$ to true or false}] at(0, -6.5){$x_n$};
    \coordinate [label=\textbf{\ldots}](elipses) at(0, -6);
    \coordinate (dummy) at (0, -5);
    \node (c1) [empty, label={[align=center, shift={(6,-1.5)}]Player Two chooses the \\ clause of the conjunctive \\ normal form forumula}] at(-3, -9.25) {$c_1$};
    \node (c2) [empty, below=of x_n] {$c_2$};
    \node (cn) [empty] at(3, -9.25) {$c_n$};
    \node (x_1new) [empty, label={[align=center, shift={(-2,-1)}]Player One chooses the\\ literal of the conjunctive \\ normal form forumula}] at (-4.75, -12) {$x_1$};
    \node (x_2new) [empty, below=of c1] {$x_2$};
    \coordinate [label=\textbf{\ldots}](clauseElipses) at (1.5, -9.85);
    \coordinate [label=\textbf{\ldots}](variableElipses) at (-1.4, -12.5);
    \node (x_nNew) [empty] at (0, -12) {$x_n$};
    \begin{pgfonlayer}{background}
        \draw[cPointed] (x_1) edge (x_2)
                        (x_2) edge (dummy)
                        (x_n) edge (c1)
                        (x_n) edge (c2)
                        (x_n) edge (cn)
                        (c1) edge (x_1new)
                        (c1) edge (x_2new)
                        (c1) edge (x_nNew);
    \end{pgfonlayer}
\end{tikzpicture}
}
\end{centering}
\caption{The general flow of play for the TQBF \ruleset{Slime Trail} game.}
\label{fig:overview}
\end{figure}

Our reduction will utilize this. We will create a legal board state of \ruleset{Slime Trail} that is an instance of QBF such that there is a winning strategy for the initial player if and only if the formula evaluates to true. This reduction is inspired by the reduction of QBF to \ruleset{Geography}\cite{PapadimitriouBook:1994}. The game will play out by allowing the the first player control the status of $x_1$, then the second player the status of $x_2$, and so on. Thus, player one sets all $x_i$ such that $i$ is odd, and player two sets all $x_i$ such that $i$ is even. At the end of the game, we "investigate" one of the variables and use its value as determined by the game to find the winner of the game. See \cref{fig:overview} for more details.

If a player breaks the general flow of play outlined above (i.e., doesn't properly set the variable) then the player will lose in a constant number of turns. Thus, each player must follow the prescribed route.

At the end of the prescribed route, player 2 selects the clause, and player 1 selects the literal. Because of this, if at least one of the clauses contains only false, then player 2 wins. Otherwise, player 1 can always select a true literal and win the game.

\subsection{Gadget Overview}

In order to reduce the QBF into an instance of the game \ruleset{Slime Trail}, we need to create gadgets to perform the actions required by the QBF game outlined in \cref{fig:overview}. In these gadgets, we assume that the first player has goal nodes colored blue and that the second player's goal nodes are colored red. We will sometimes refer to the players as "Blue" or "Red" respectively, for convenience.

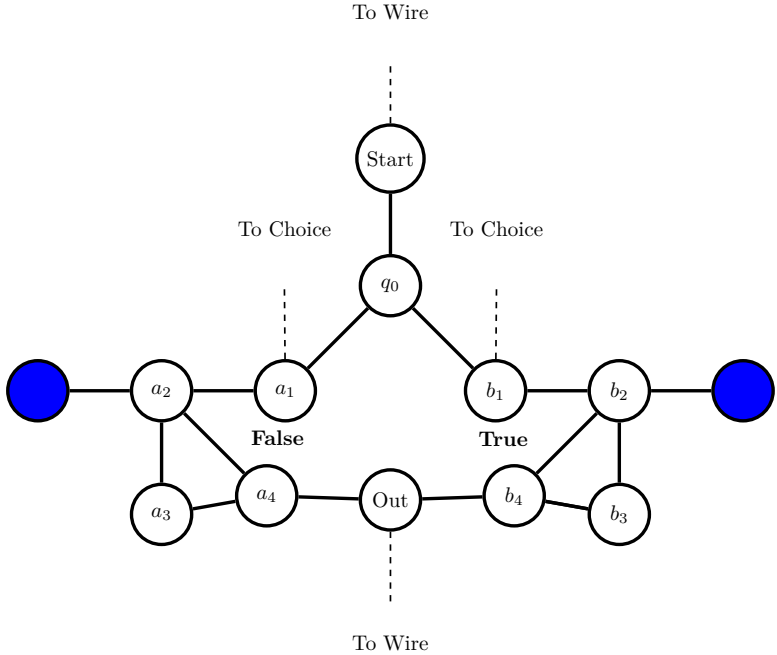
\begin{figure}[h!]
\begin{centering}
\scalebox{.8}{%
\begin{tikzpicture}[scale=.75, every node/.style={draw,ultra thick,circle,minimum size = .3cm} ]
    \node (s) [empty] {$q_0$};
    \node (s2) [empty, above=of s] {Start};
    \node (a1) [empty, below left=of s] {$a_1$};
    \node (b1) [empty, below right=of s] {$b_1$};
    \node (a2) [empty, left=of a1] {$a_2$};
    \node (b2) [empty, right=of b1] {$b_2$};
    \node (blue1) [blue, left=of a2] {};
    \node (blue2) [blue, right=of b2] {};
    \node (a3) [empty, below=of a2] {$a_3$};
    \node (a4) [empty, below right=of a2] {$a_4$};
    \node (b3) [empty, below=of b2] {$b_3$};
    \node (b4) [empty, below left=of b2] {$b_4$};
    \node (out) [empty] at(0, -4.75) {Out};
    \coordinate [label=To Choice](c1) at(-2.35, 0);
    \coordinate [label=To Choice](c2) at(2.35, 0);
    \coordinate [label=To Wire](c3) at(0, 5);
    \coordinate [](c4) at(0, -7);
    \coordinate [label=To Wire](c4B) at(0, -9);
    \coordinate [label=\textbf{True}](t) at(2.5, -4.2);
    \coordinate [label=\textbf{False}](f) at(-2.5, -4.2);

    \begin{pgfonlayer}{background}
        \draw[cLine] (s) edge (a1)
                     (s) edge (b1)
                     (s2) edge (s)
                     (a1) edge (a2)
                     (b1) edge (b2)
                     (a2) edge (a3)
                     (a2) edge (blue1)
                     (a2) edge (a4)
                     (b2) edge (b3)
                     (b2) edge (b4)
                     (b2) edge (blue2)
                     (a3) edge (a4)
                     (a4) edge (out)
                     (b3) edge (b4)
                     (b3) edge (b4)
                     (b4) edge (out);

        \draw[cDashed]  (a1) edge (c1)
                        (b1) edge (c2)
                        (s2) edge (c3)
                        (out) edge (c4);
    \end{pgfonlayer}
\end{tikzpicture}
}
\end{centering}
\caption{The even variables. The starting position is Start, and it is Red's turn to move.}
\label{fig:even}
\end{figure}

In order to replicate the QBF game, we will need:

\begin{itemize}
\item An odd variable gadget (\cref{fig:odd}) for the first player to assign values
\item An even variable gadget (\cref{fig:even}) for the second player to assign values
\item A choice gadget (\cref{fig:choice}) for Player Two to select the clause and for Player One to select the literal.
\end{itemize}
Additionally, in order to resolve some issues created within the gadgets, we have the following "helper" gadgets.
\begin{itemize}
\item A wire gadget (\cref{fig:wire}) to ensure that, regardless of who moved last on the previous variable, it is Blue's turn at the start of the next gadget.
\item A crossover gadget (\cref{fig:crossover}) to ensure that the reduction holds even when restricted to a planar graph.s
\end{itemize}

Each of these gadgets are discussed in detail in their own section.

\subsection{Odd Variable Gadget}

\begin{figure}[h!]
\begin{centering}
\scalebox{.8}{%
\begin{tikzpicture}[scale=.75, every node/.style={draw,ultra thick,circle,minimum size = .3cm} ]
    \node (s) [empty] {Start};
    \node (a1) [empty, below left=of s] {$a_1$};
    \node (b1) [empty, below right=of s] {$b_1$};
    \node (a2) [empty, left=of a1] {$a_2$};
    \node (b2) [empty, right=of b1] {$b_2$};
    \node (a3) [empty, left=of a2] {$a_3$};
    \node (b3) [empty, right=of b2] {$b_3$};
    \node (blue1) [blue, above=of a3] {};
    \node (blue2) [blue, above=of b3] {};
    \node (a4) [empty, below=of a3] {$a_4$};
    \node (a5) [empty, below right=of a3] {$a_5$};
    \node (b4) [empty, below=of b3] {$b_4$};
    \node (b5) [empty, below left=of b3] {$b_5$};
    \node (out) [empty] at(0, -4.75) {Out};
    \coordinate [label=To Choice](c1) at(-5.15, 0);
    \coordinate [label=To Choice](c2) at(5.15, 0);
    \coordinate [](c4) at(0, -7);
    \coordinate [label=To Wire](c4B) at(0, -9);
    \coordinate [label=\textbf{False}](f) at(-4, -4.5);
    \coordinate [label=\textbf{True}](t) at(4, -4.5);

    \begin{pgfonlayer}{background}
        \draw[cLine] (s) edge (a1)
                     (s) edge (b1)
                     (a1) edge (a2)
                     (b1) edge (b2)
                     (a2) edge (a3)
                     (b2) edge (b3)
                     (a3) edge (blue1)
                     (a3) edge (a4)
                     (a3) edge (a5)
                     (b3) edge (blue2)
                     (b3) edge (b4)
                     (b3) edge (b5)
                     (a4) edge (a5)
                     (b4) edge (b5)
                     (a5) edge (out)
                     (b5) edge (out);

        \draw[cDashed]  (a2) edge (c1)
                        (b2) edge (c2)
                        (out) edge (c4);
    \end{pgfonlayer}
\end{tikzpicture}
}
\end{centering}
\caption{The odd variables. At Start, it is blue's turn to move.}
\label{fig:odd}
\end{figure}
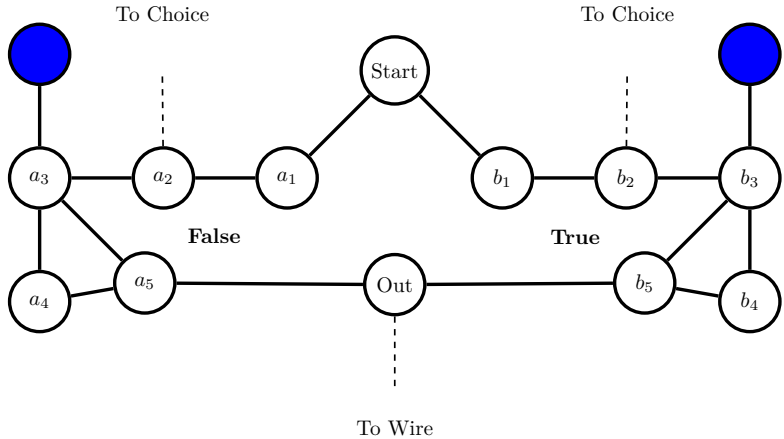

In the odd variable gadget (described in \cref{fig:odd}), we start at "Start," where it is the Blue player's turn. If this isn't the first variable, then Red just moved to Start from the wire gadget (\cref{fig:wire}). Blue can choose to either move to $a_1$(equivalent to setting the associated variable to false), or to $b_1$ (equivalent to setting the associated variable to true). For the purpose of this walk through, we presume that Blue moves left, but they could also move right. Since both sides are symmetric, it plays similarly.

After moving, it is Red's turn, who must move to $a_2$. Blue now can either move to $a_3$, or to the choice gadget (\cref{fig:choice}). If they go the choice gadget, Red has an immediate trivial winning move (It is also possible that it connects to a crossover gadget\cref{fig:crossover}. In this case, Red will again have a winning move). Because moving to the choice gadget will result in a loss, Blue should choose to move toward $a_3$. Now that it is Red's turn again, they can choose to immediately lose, move to $a_4$, or move to $a_5$. Either is fine.

If they move to $a_4$, then Blue must move to $a_5$. Red must then move to Out. Here, Blue can choose to either move down to the wire gadget (\cref{fig:wire}), or Blue can move right to $b_5$. Moving to $b_5$ is a losing move for Blue, since Red can move to $b_4$, which forces Blue to move to $b_3$, which causes Red to move to $b_2$. Now, it is Blue's move, and $b_1$ is not an option because the rules specify that there must always be a path to at least one of the goal nodes, so Blue must move to the choice gadget. Since, as previously mentioned, this results in an instantaneous loss for Blue, Blue loses. Thus, if Red chooses to move to $a_4$ from $a_3$, then then Blue should choose to move from Out to the wire gadget \cref{fig:wire}.

If Red chooses to move to $a_5$ from $a_3$, then Blue must move to Out, since trapping the slime at $a_4$ violates the rules. Now it is Red's move. They can either move to the wire gadget, or move to $b_5$. If they move to $b_5$, they will lose, as Blue will move to $b_4$, forcing Red to $b_3$. Blue then immediately wins by moving to the Blue goal node.

So, if both players are playing optimally, they will move from Out to the wire gadget below it (\cref{fig:wire}). It can end on either Blue or Red's move, depending on whether Red chooses $a_4$ or $a_5$. We will end this gadget with one of the sides "slimed" so that it is inaccessible from the Choice gadget. The other side will still be reachable, which is important for the Choice gadget ({\cref{fig:choice}}) used at the very end of the game.

\subsection{Even Variable Gadget}

In this gadget (as seen in \cref{fig:even}), we start at "Start," where it is again the Blue player's turn, who must move to $q_0$. Now, Red can choose to either move left to $a_1$ (setting the associated variable to false), or right (setting the associated variable to true). For the purpose of this walk through, we are going to presume that Red moved left, but it could have also moved right. Since both sides are symmetric, it plays similarly.

After moving, Blue can either move to $a_3$, or to the choice gadget (\cref{fig:choice}). If they go the choice gadget, Red has an immediate trivial winning move (It is also possible that it connects to a crossover gadget, as seen in \cref{fig:crossover}. In this case, Red will again have a winning move). Because moving to the choice gadget will result in a loss, Blue should choose to move toward $a_2$. Now that it is Red's turn again, they can choose to immediately lose, move to $a_3$, or move to $a_4$. This plays out similarly to the Odd Variable gadget.

\begin{figure}[h!]
\begin{centering}
\scalebox{.7}{%
\begin{tikzpicture}[scale=.75, every node/.style={draw,ultra thick,circle,minimum size = .3cm} ]
    \node (s) [empty] {Start};
    \node (a1) [empty, below left=of s] {$a_1$};
    \node (b1) [empty, below right=of s] {$b_1$};
    \node (a2) [empty, left=of a1] {$a_2$};
    \node (b2) [empty, below=of b1] {$b_2$};
    \node (blue1) [blue, above left=of a2] {};
    \node (red1) [red, below left=of a2]{};
    \node (b3) [empty, right=of b2] {$b_3$};
    \node (blue2) [blue, above right=of b3] {};
    \node (red2) [red, below right=of b3] {};
    \node (a3) [empty, below=of a1] {$a_3$};
    \node (a4) [empty, below=of a3] {$a_4$};
    \node (blue3) [blue, left=of a4] {};
    \node (b4) [empty, below=of b2] {$b_4$};
    \node (red3) [red, right=of b4] {};
    \node (a5) [empty, below=of a4] {$a_5$};
    \node (out) [empty] at(0, -12) {Out};
    \coordinate [](c1) at(0, 3);
    \coordinate [label=From Variable](c1B) at(0, 2);
    \coordinate [](c2) at(0, -14);
    \coordinate [label=To Variable](c2B) at(0, -16);

    \begin{pgfonlayer}{background}
        \draw[cLine] (s) edge (a1)
                     (s) edge (b1)
                     (a1) edge (a2)
                     (a1) edge (a3)
                     (b1) edge (b2)
                     (a2) edge (blue1)
                     (a2) edge (red1)
                     (b2) edge (b3)
                     (b2) edge (b4)
                     (a3) edge (a4)
                     (a4) edge (a5)
                     (a4) edge (blue3)
                     (a5) edge (out)
                     (a5) edge (out)
                     (b3) edge (blue2)
                     (b3) edge (red2)
                     (b4) edge (out)
                     (b4) edge (red3);

       \draw[cDashed]   (c1) edge (s)
                        (c2) edge (out);
    \end{pgfonlayer}
\end{tikzpicture}
}
\end{centering}
\caption{The wire gadget, witch forces red to play from "out". The starting position is Start, and it is either red or blue's turn to move.}
\label{fig:wire}
\end{figure}
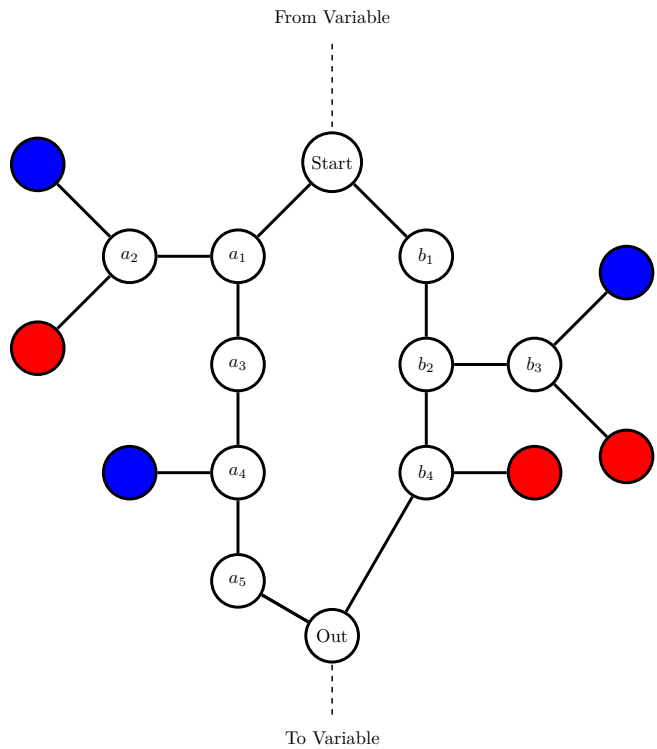

If they move to $a_3$, then Blue must move to $a_4$. Red must then move to Out. Here, Blue should move on to the wire gadget (\cref{fig:wire}). The other option is to move right to $b_4$, which is a losing move for Blue. Red would move to $b_3$, which forces Blue to move to $b_2$, which causes Red to move to $b_1$. Now, it is Blue's move, and they must move to the choice gadget. Since, as previously mentioned, this results in an instantaneous loss for Blue, Blue loses. Thus, if Red chooses to move to $a_3$, then the game should proceed with Blue moving to the wire gadget (\cref{fig:wire}).

If Red chooses to move to $a_2$ from $a_4$, then Blue must move to Out, as trapping the token at $a_3$ violates the rules of the game. Now it is Red's move. They can either move to the wire gadget, or move to $b_4$. If they move to $b_4$, they will lose, as Blue will move to $b_3$, forcing Red to $b_2$. Blue then immediately wins by moving to the goal node.

So, if both players are playing optimally, they will move from Out to the wire gadget below it (\cref{fig:wire}). It can end on either Blue or Red's move. Like in the odd variable gadget, we will end this gadget with one of the sides "slimed" so that it is inaccessible from the Choice gadget. The other side will still be reachable, which is important for the Choice gadget ({\cref{fig:choice}}) at the very end of the game.

\subsection{Wire}

\begin{figure}[h!]
\begin{centering}
\scalebox{.6}{%
\begin{tikzpicture}[scale=.75, minimum size = .3cm]
    \node (s) [empty] {$c_0$};
    \node (s2) [empty, above=of s] {Start};
    \node (c1) [empty] at(-10.5, -2.87) {$c_1$};
    \node (c2) [empty, below=of s] {$c_2$};
    \node (cn) [empty] at(10.5, -2.87) {$c_n$};
    \node (red1) [red, left=of c1] {};
    \node (red2) [red, left=of c2] {};
    \node (redN) [red, left=of cn] {};
    \coordinate (v1) at(-14, -6);
    \coordinate (v12) at(-10.5, -6);
    \coordinate (v1N) at(-7, -6);
    \coordinate (v2) at(-3.5, -6);
    \coordinate (v22) at(0, -6);
    \coordinate (v2N) at(3.5, -6);
    \coordinate (vN) at(7, -6);
    \coordinate (vN2) at(10.5, -6);
    \coordinate (vNN) at(14, -6);
    \coordinate [label=To Variable $a_1$](v1B) at(-14, -7);
    \coordinate [label=To Variable $a_2$](v12B) at(-10.5, -7);
    \coordinate [label=To Variable $a_N$](v1NB) at(-7, -7);
    \coordinate [label=To Variable $b_1$](v2B) at(-3.5, -7);
    \coordinate [label=To Variable $b_2$](v22B) at(0, -7);
    \coordinate [label=To Variable $b_N$](v2NB) at(3.5, -7);
    \coordinate [label=To Variable $n_1$](vNB) at(7, -7);
    \coordinate [label=To Variable $n_2$](vN2B) at(10.5, -7);
    \coordinate [label=To Variable $n_N$](vNNB) at(14, -7);
    \coordinate (w) at (0, 5);
    \coordinate [label=From Wire](wB) at (0, 6);
    \node at (4, -2.9) {\textbf{\ldots}};
    \node at (-9, -5.7) {\textbf{\ldots}};
    \node at (1.5, -5.7) {\textbf{\ldots}};
    \node at (12, -5.7) {\textbf{\ldots}};

    \begin{pgfonlayer}{background}
        \draw[cLine] (s) edge (s2)
                     (s) edge (c1)
                     (s) edge (c2)
                     (s) edge (cn)
                     (c1) edge (red1)
                     (c2) edge (red2)
                     (cn) edge (redN);
        \draw[cDashed] (c1) edge (v1)
                       (c1) edge (v12)
                       (c1) edge (v1N)
                       (c2) edge (v2)
                       (c2) edge (v22)
                       (c2) edge (v2N)
                       (cn) edge (vN)
                       (cn) edge (vN2)
                       (cn) edge (vNN)
                       (s2) edge (w);
    \end{pgfonlayer}
\end{tikzpicture}
}
\end{centering}
\caption{The choice gadget. The starting position is Start, and it is red's turn to move.}
\label{fig:choice}
\end{figure}
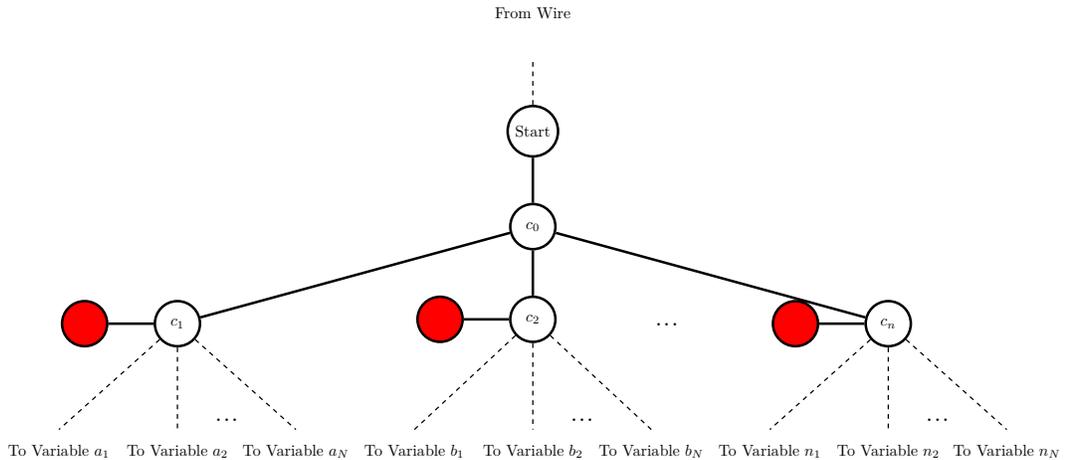

The wire gadget is described in \cref{fig:wire}. Here, we begin at Start. We have just arrived from one of the variable gadgets, and the current player is either Red or Blue. Our goal is to make it so that Red moves from Out to the Start node of one of the variable gadgets, or to the Start node of the choice gadget (\cref{fig:choice}).

If it is Blue's turn, they can either move to $a_1$ or $b_1$. However, moving to $b_1$ is a losing move. If they do, Red moves to $b_2$, which leaves Blue with no options that don't result in a Red victory one turn later.

If Blue chooses $a_1$. Red can then move to $a_2$ or $a_3$. It is never in any player's interest to move to $a_2$, as the other player always has an immediate winning move. Thus, Red should choose $a_3$. Blue must move to $a_4$, which Red should follow with moving to $a_5$. Blue then moves to Out. Red can then either make a move to the next gadget, or move to $b_3$. Moving to $b_3$ is a losing move, since Blue can move to $b_2$, forcing Red to make the losing move to $b_3$, which functions similarly to $a_2$.

If it is Red's turn, they can either move to $a_1$ or $b_1$. $a_1$ is a losing move, as Blue can move to $a_3$, which forces Red to move to $a_4$ and lose. Thus, Red should choose $b_1$. Blue must move to $b_2$, from which Red should choose $b_4$, since $b_3$ is an undesirable move. Blue should move to Out, again leaving Red the choice of moving to the next gadget, or moving to $a_5$. Moving to $a_5$ is a losing move, as play alternates until the players reach $a_1$, where Red is forced to move to $a_2$ and Blue wins the game.

Thus, the gadget successfully makes it Red's move to play on Out, with any available moves not from Out into the next gadget causing Red to lose in a constant number of moves.

\subsection{Choice}
The choice gadget is described in \cref{fig:choice}. We begin with Red moving from the Wire gadget into Start. From here, Blue must move to $c_0$. Red now chooses which clause Blue must play from. Each clause will have some literals for the Blue player to pick from.

These literals were all set by the two players previously in the variable gadgets when the players chose to move to either left or right, setting the variable to true or false. If the literal in the expression is $x_1$, for example, the "to variable" will lead to the "to choice" on the right of the first variable set. If Blue slimed that area, they will be unable to make that move. If all of the literals are unavailable because of the choices made in the game above, then Blue's only move is to move into Red's goal node.

\begin{figure}[h!]
\begin{centering}
\scalebox{.7}{%
\begin{tikzpicture}[scale=.75, minimum size = .3cm]
    \node (s1) [empty] {$Start_a$};
    \node (s2) [empty] at(7, 8.35) {$Start_b$};
    \node (a1) [empty, above=of s] {$a_1$};
    \node (a2) [empty, above=of a1] {$a_2$};
    \node (blue1) [blue, left=of a2] {};
    \node (b1) [empty, left=of s2] {$b_1$};
    \node (red1) [red, above=of b1] {};
    \node (cross) [empty, above=of a2] {Cross};
    \node (a3) [empty, above=of cross] {$a_3$};
    \node (a4) [empty, above=of a3] {$a_4$};
    \node (blue3) [blue, left=of a3] {};
    \node (b2) [empty, left=of cross] {$b_2$};
    \node (red2) [red, below left=of b2] {};
    \node (red3) [red, left=of a4] {};
    \coordinate [label=To Variable](v1) at(-8, 8.35);
    \coordinate [label=To Variable](v2) at (0, 16);
    \coordinate [](c1) at (0, -2.5);
    \coordinate [label=From Choice](c1B) at (0, -3.5);
    \coordinate [label=From Choice](c2) at (11, 8.35);

    \begin{pgfonlayer}{background}
        \draw[cLine] (s) edge (a1)
                     (a1) edge (a2)
                     (a2) edge (blue1)
                     (a2) edge (cross)
                     (s2) edge (b1)
                     (b1) edge (cross)
                     (b1) edge (red1)
                     (cross) edge (a3)
                     (cross) edge (b2)
                     (a3) edge (blue3)
                     (a3) edge (a4)
                     (a4) edge (red3)
                     (b2) edge (red2);
      \draw[cDashed] (b2) edge (v1)
                     (a4) edge (v2)
                     (s1) edge (c1)
                     (s2) edge (c2);
    \end{pgfonlayer}
\end{tikzpicture}
}
\end{centering}
\caption{The crossover gadget. The starting position is either $Start_a$ or $Start_b$, and it is red's turn to move.}
\label{fig:crossover}
\end{figure}
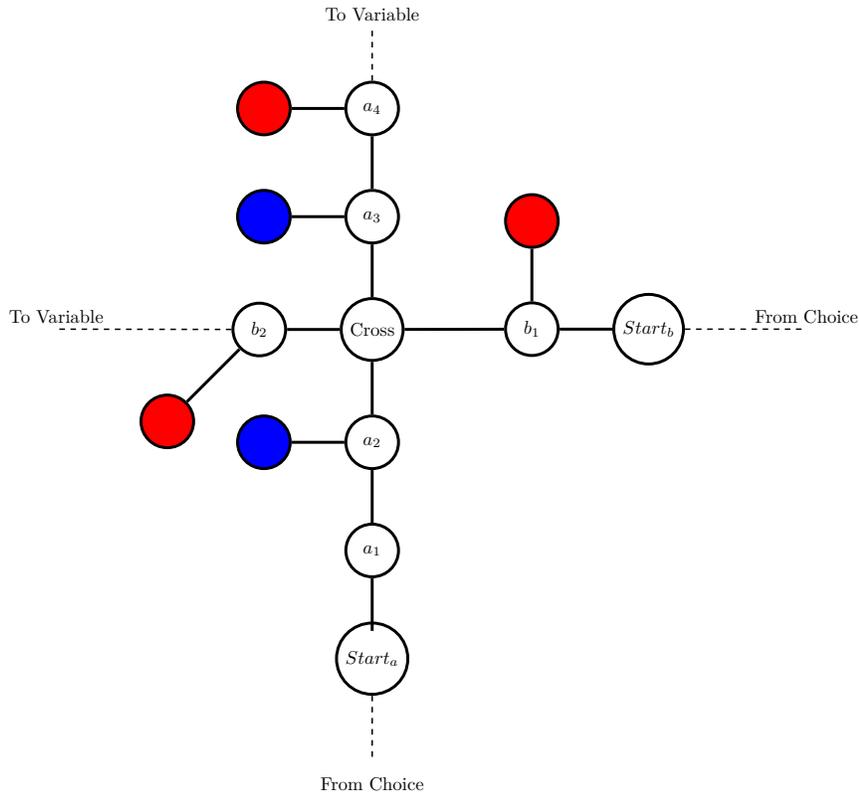

\subsection{Crossover}
The crossover gadget is described in \cref{fig:crossover}. The crossover gadget exists to ensure that the reduction applies even on planar graphs. In the locations where the choice gadget leads back to the variables, the edges could intersect. To resolve this issue, we have the crossover gadget. The player starts from either $Start_a$ or $Start_b$. If they begin in $Start_a$, it should end on $a_4$, and if the player begins in $Start_b$, then they must end on $b_2$. It is always Red's move at the start.

If starting from $Start_a$, Red must move first to $a_1$. From there, Blue must move to $a_2$. To avoid losing, Red must move to $Cross$. If the Blue player moves to either $b_1$ or $b_2$, they lose the game, so they must move to $a_3$. Red must move to $a_4$ to avoid losing, and then Blue moves to the variable as expected before.

If starting from b, Red must move to $b_1$. Blue must then move to $Cross$. Red will lose in one move if they move to $a_2$ or $a_3$, so they must move to $b_2$. From here, Blue continues to the variable.

\subsection{Conclusion}
Using the gadgets we created, we can create the flow of play mentioned in \cref{fig:overview}. We have odd variables for the first player to set to true or false, and even variables for the second player to set to true or false. We have a choice gadget to allow the second player to choose a clause which includes multiple literals and player one to choose a literal from that clause. We have a wire gadget to ensure uniformity between the gadgets, since all other gadgets are now able to begin on Blue's turn. We have a crossover gadget to ensure that the game is played on a planar graph. With the flow of play now observed, we have successfully made a valid board of \ruleset{Slime Trail} from an instance of QBF. Thus, \ruleset{Slime Trail} is PSPACE-hard on a planar graph. Combined with Lemma 3.2, we now know that \ruleset{Slime Trail} is PSPACE-complete when played on a planar graph.

\section{Future Research}
\label{sec: future}
\ruleset{Slime Trail} is typically played on a square or hexagonal grid. This leads us to the question:
\begin{open}[\ruleset{Slime Trail} on a Grid]
Is \ruleset{Slime Trail} still PSPACE-complete when played on a grid?
\end{open}
The game is often played with only one goal node per player. Since our reduction relies on multiple goal nodes, we also have another question:
\begin{open}[One goal node \ruleset{Slime Trail}]
What is the tractability of \ruleset{Slime Trail} when each player has only one available goal vertex?
\end{open}

\bibliographystyle{plain}

\end{document}